\documentclass[superscriptaddress,showpacs,a4paper]{revtex4}
\usepackage{subeqn}
\usepackage{graphicx}
\usepackage{numprint}
\usepackage[latin1]{inputenc}

\newcommand{\EVRY}{Universit\'e d'Evry-Val d'Essonne, Boulevard François Mitterrand, F-91000 Evry, France}
\newcommand{\LKB}{Laboratoire Kastler Brossel, UPMC-Sorbonne Universit\'es, CNRS, ENS-PSL Research University, Coll\`ege de France, 4 place Jussieu, F-75005 Paris, France}
\newcommand{\bN}{\mbox{\boldmath$\nabla$}}
\newcommand{\br}{\mathbf{r}}
\npdecimalsign{.}

\begin{document}
\title{Resonances in two-electron atoms below the critical charge}

\author{Jean-Philippe Karr}
\email{karr@lkb.upmc.fr}
\affiliation{\LKB}
\affiliation{\EVRY}

\date{\today}
\begin{abstract}
The critical nuclear charge $Z_c$ required for a heliumlike atom to have at least one bound state was recently determined with high accuracy from variational calculations. Analysis of the wave functions further suggested that the bound state changes smoothly into a shape resonance as $Z$ crosses the critical value. Using variational calculations combined with the complex coordinate rotation method, we study the energy and width of the resonance for $Z < Z_c$, thus providing direct evidence of the validity of this hypothesis. The variation of the resonance width with $Z$ is found to be in good agreement with a model derived from analysis of the $1/Z$ perturbation series.
\end{abstract}
\maketitle

\section{Introduction}

Let us consider the general problem of a two-electron atom with a nuclear charge $Z$ and infinite nuclear mass, described, in atomic units, by the Hamiltonian
\begin{equation}
H_0 = -\frac{\bN_{\br_1}^2}{2} - \frac{\bN_{\br_2}^2}{2} -\frac{Z}{r_1} - \frac{Z}{r_2} +\frac{1}{r_{12}},
\end{equation}
where $\br_i$ is the position of electron $i$ with respect to the nucleus and $r_{12}$ the inter-electron distance. The scaling transformation $\br_i \rightarrow \br_i/Z$, $E \rightarrow E/Z^2$ yields
\begin{equation}
H = -\frac{\bN_{\br_1}^2}{2} - \frac{\bN_{\br_2}^2}{2} -\frac{1}{r_1} - \frac{1}{r_2} +\frac{1}{Z} \frac{1}{r_{12}}. \label{ham}
\end{equation}
It is then possible to treat the electron-electron interaction as a perturbation, with a perturbation parameter $\lambda = 1/Z$, leading to the following expansion for the ground-state energy $E(Z)$:
\begin{equation}
E(Z) = \sum_{n=0}^{\infty} \frac{E_n}{Z^n}. \label{series}
\end{equation}
This perturbation series attracted considerable interest, being one of the few convergent perturbation expansions in atomic and molecular physics. A long-standing goal has been to determine the radius of convergence $\lambda^{\ast} = 1/Z^{\ast}$ as precisely as possible, and understand the relationship between $Z^{\ast}$ and the critical charge $Z_c$, defined as the charge for which the ground state energy reaches the first detachment threshold, i.e. $E(Z_c) = -1/2$. An important step forward was the calculation of the first 402 coefficients of the perturbation expansion by Baker {\textit et al.}~\cite{baker1990}, whose subsequent analysis yielded the estimates $Z^{\ast} \simeq \numprint{0.91103}$~\cite{baker1990} and $Z^{\ast} \simeq \numprint{0.91102826}$~\cite{ivanov1995}. On the other hand, a recent study of $E(Z)$ using high-precision variational calculations~\cite{estienne2014} yielded $Z_c = \numprint{0.91102822407725573}(4)$, later confirmed in~\cite{pilon2015}, implying that $Z^{\ast} = Z_{c}$ within the current accuracy of the determination of $Z^{\ast}$.

However, the nature of the transition occurring at the critical point has not been fully elucidated yet. Early studies by Stillinger~\cite{stillinger1966,stillinger1974} as well as more recent ones~\cite{turbiner2015} connected with doubts on the accuracy of the determination of $Z^{\ast}$~\cite{turbiner2014} suggested that the system could persist as a bound state embedded in the continuum even for $Z<Z_c$. This possibility was questioned by Reinhardt in an analysis based on the theory of dilation analyticity~\cite{reinhardt1977}. In~\cite{estienne2014} an analysis of the variational wave functions showed that the outer electron remains localized near the nucleus even for $Z \simeq Z_c$, and a simple model of the potential it experiences pointed to the existence of shape resonances. Overall the results of~\cite{estienne2014} are consistent with a smooth transition from a bound state for $Z > Z_c$ to a shape resonance for $Z < Z_c$.

In this context, it is highly desirable to investigate numerically the existence and properties of resonances for $Z < Z_c$. Resonances were studied with the complex rotation method in~\cite{dubau1998}, but in that work only the region $\lambda \geq 1.11$ (i.e. $Z \leq 0.9009$) was explored, still relatively far from the critical value. Here we present more accurate and complete results where the range of investigation was extended up to $Z = 0.9103$. Our results are in full agreement the hypothesis of a smooth transition from a bound state to a resonance formulated in~\cite{estienne2014}, and put much stronger constraints on the possible existence of a bound state for $Z < Z_c$. Furthermore, a model of the resonance width derived solely from the analysis of the asymptotic form of the $1/Z$ perturbation coefficients~\cite{ivanov1998} is in good agreement with our data, and gives further indication that the state of the system starts acquiring a finite width exactly at $Z = Z_c$.

\section{Numerical calculation of resonances}

In order to obtain the energies and widths of resonances, we diagonalize numerically the complex rotated Hamiltonian~(\ref{ham}) in a variational basis set. We use the perimetric coordinates (denoted $x,y,z$) and Sturmian wave functions
\begin{equation}
 \chi_{n_x,n_y,n_z}^{\alpha,\beta} (x,y,z) = (-1)^{n_x + n_y + n_z} \sqrt{\alpha\beta^2} \times L_{n_x}\left(\alpha x\right) L_{n_y} \left(\beta y\right) L_{n_z} \left(\beta z\right) e^{-(\alpha x+\beta y+\beta z)/2}
\end{equation}
where $n_x,n_y,n_z$ are non-negative integers and $L_{n}$ the Laguerre polynomials~\cite{pekeris1958}. The variational basis is truncated by imposing the conditions
\begin{equation}
n_x + n_y + n_z \leq n^{max} \; , \;\; n_x \leq n_x^{max}.
\end{equation}

This approach allows to write the Hamiltonian in the form of a sparse-band matrix, and gives high numerical accuracy with double-precision arithmetic~\cite{hilico2000,karr2006}. It has been shown to yield very precise results for the complex energy of resonances~\cite{kilic2004}. Finally it is well suited for studies where parameters of the physical system are changed (such as, here, the nuclear charge $Z$). Indeed, the small number of variational parameters (the two length scales $\alpha^{-1}$ and $\beta^{-1}$) makes it easy to readjust them for each value of $Z$.

\begin{figure}[h]
\begin{center}
\includegraphics[width=10cm]{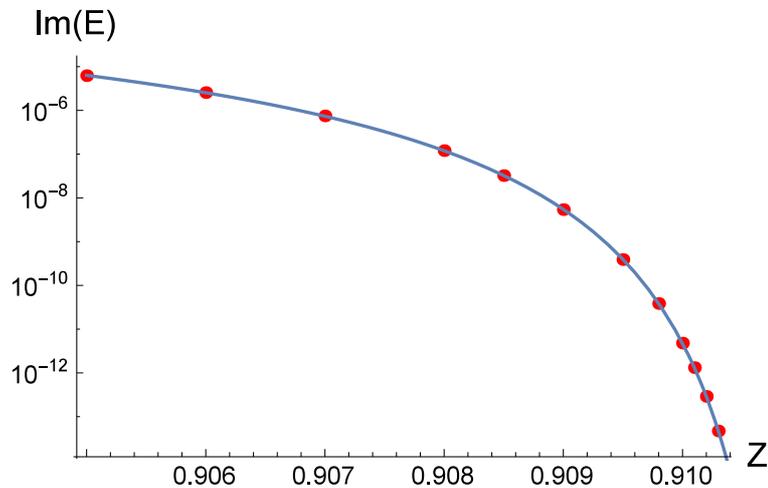}
\end{center}
\caption{Imaginary part of the energy of the lowest resonance of a two-electron atom as a function of the nuclear charge $Z$. The solid line is a fit by expression~(\ref{fit}) (see text for details).}
\label{curve}
\end{figure}

Our results are summarized in Table~\ref{results} and Figure~\ref{curve}. We cross-checked the determination of $Z_c$ in Refs.~\cite{estienne2014,pilon2015} by computing the energy for $Z =\numprint{0.9110282440772}$ and $Z = \numprint{0.9110282440773}$ (see the last two lines of Table~\ref{results}). In both cases all 15 digits are in agreement with those of~\cite{estienne2014}. We thus confirm that $Z_c$ lies between these two values; a more detailed analysis is possible but would lie outside the scope of the present work. This also shows that our method can reach an accuracy of about $1 \times 10^{-15}$ a.u. that is essentially limited by numerical noise.

In the region $Z < Z_c$, we checked the closest value to $Z_c$ studied so far~\cite{dubau1998}, i.e. $Z = 1/1.11 = \numprint{0.9009009}...$ and find
\begin{equation}
{\rm Re}(E) = \numprint{-0.49713121236759} \, \mbox{,} \;\; {\rm Im}(E) = \numprint{-4.997292209} \, 10^{-5}.
\end{equation}
This is in agreement with, and much more accurate than the result of~\cite{dubau1998}, ${\rm Im}(E) = -6(3) \, 10^{-5}$. The study was then pursued by increasing $Z$ in small steps, going as close to $Z = Z_c$ as possible. As can be seen from Table~\ref{results} and Figure~\ref{curve}, the width of the resonance drops rapidly with increasing $Z$. The $\sim 10^{-15}$ a.u. numerical accuracy of the calculation implies that only resonance widths in the $10^{-14}$ a.u. range or larger can be obtained with sufficient accuracy. This is what limits the explored range to $Z = 0.9103$ where the width is $4.7(1) \, 10^{-14}$. At $Z = 0.9104$ the width is already too small to be resolved.

Overall, our results confirm that the bound states is transformed into a resonance. In addition, the parameters $\alpha,\beta$ of the variational wave function are found to vary smoothly as a function of $Z$, in agreement with what was observed in~\cite{estienne2014}, suggesting a smooth transition from bound to resonant state. Rigorously speaking, there is no proof that this transition occurs exactly at $Z = Z_c$, since the interval $Z \in [0.9103,Z_c]$ was not studied. One could improve further the accuracy of the calculation by using quadruple precision arithmetic and a larger variational basis. However, in view of the extremely fast decrease of ${\rm Im}(E)$, this would not allow to get much closer to $Z_c$ and would not be worth the increased computational effort. Whatever the accuracy of the calculation, some small range in the vicinity of the critical point would remain beyond reach. Thus, the only clear-cut conclusion is that
\begin{equation}
Z^{\ast} > 0.9103
\end{equation}

However, there is no physical argument to support the hypothesis that, for decreasing $Z$, a bound state would persist for some small range $[Z^{\ast},Z_c]$ and then disappear at $Z = Z^{\ast}$. As discussed in~\cite{reinhardt1977,stillinger1977}, such a scenario is compatible with the theory of dilation analyticity only if this bound state is not square-integrable. This possibility seems to be ruled out by the observed smooth variation of the parameters describing the wave function.

In order to gain further insight on the nature of the atomic state in the range $Z \in [0.9103,Z_c]$ and estimate the value of $Z$ at which the width of the resonance goes to zero, we will use a model of the resonance width derived by Ivanov and Dubau~\cite{ivanov1998}. This is the object of the next Section.

\begin{table}
\begin{tabular}{n{1}{14}n{1}{15}n{1}{8}l}
\hline
{$Z$} & {Re($E$)} & {Im($E$)} \\
\hline
0.905	& -0.49826110390007	 & -6.22402727 & $10^{-06}$ \\
0.906 	& -0.49854250216297	 & -2.55236730 & $10^{-06}$ \\
0.907	& -0.49882687942493	 & -7.4824776  & $10^{-07}$ \\
0.908	& -0.49911431212279	 & -1.2038420  & $10^{-07}$ \\
0.9085	& -0.49925911100848	 & -3.233035   & $10^{-08}$ \\
0.909	& -0.49940455976678	 & -5.39792    & $10^{-09}$ \\
0.9095	& -0.49955059334234	 & -3.8935     & $10^{-10}$ \\
0.9098	& -0.49963846869642  & -3.850      & $10^{-11}$ \\
0.910	& -0.49969715161159	 & -4.79       & $10^{-12}$ \\
0.9101	& -0.49972652179942	 & -1.32       & $10^{-12}$ \\
0.9102	& -0.49975591071112	 & -2.9        & $10^{-13}$ \\
0.9103	& -0.499785318046200 {(1)} & \npmakebox{-4.7 {(1)}} & $10^{-14}$ \\
\hline
0.9110282240772 & -0.499999999999983 {(1)} & {{-}} \\
0.9110282240773 & -0.500000000000013 {(1)} & {{-}} \\
\hline
\end{tabular}
\caption{Real and imaginary part of the energy of the lowest resonance of a two-electron atom of nuclear charge $Z$ (in atomic units). Typical values of the parameters are: $n^{max} = 160$, $n_x^{max} = 25$ (yielding a basis size $N = \numprint{145639}$), $\alpha \sim 1.5-3$, $\beta \sim 0.4-0.8$, complex rotation angle $\theta \sim 0.3-0.6$. All digits are significant unless otherwise noted. A larger basis set (up to $n^{max} = 200$, $n_x^{max} = 33$, yielding $N = \numprint{293301}$) was used for the last three lines in order to get one more significant digit. \label{results}}
\end{table}

\section{Model of resonance width}

Let us briefly summarize the approach of Ivanov and Dubau~\cite{ivanov1998}. They first derived a dispersion relation connecting the real and imaginary parts of the $E(Z)$ function, relying on the following analytic properties:\\
(i) $E(Z)$ is a regular function for $Z > Z^{\ast}$ (since the series~(\ref{series}) is convergent). \\
(ii) It has an essential singularity at the point $Z^{\ast}$. An indication of this behavior was obtained in~\cite{baker1990} from approximating $E(\lambda)$ by a known function whose Taylor coefficients match very closely the perturbation coefficients.\\
(iii) It has a second-order pole at $Z_0 = 0$, as shown in~\cite{ivanov1996}. \\
(iv) It has a third singular point on the real axis for some value $Z_2$, which was estimated to $Z_2 \sim 0.11$ in~\cite{ivanov1996}.\\
The dispersion relation reads
\begin{equation}
E(Z) = E_0 + \frac{C_{-1}}{Z} + \frac{C_{-2}}{Z^2} + \frac{1}{\pi} \lim_{\epsilon \to 0^+} \int_{Z_2}^{Z^{\ast}} \frac{\mbox{Im} E(t - i\epsilon)}{Z-t} \, dt,
\end{equation}
where $C_{-1}$ and $C_{-2}$ are real coefficients. From this expression one can deduce the following formula for the coefficients of the expansion~(\ref{series}):
\begin{equation}
E_n = \frac{1}{\pi} \int_{Z_2}^{Z^{\ast}} \lim_{\epsilon \to 0^+}  \mbox{Im} E(t - i\epsilon) \, t^{n-1} \, dt
\end{equation}
The following asymptotic large-$n$ behavior of the coefficients $E_n$ was established in~\cite{baker1990}:
\begin{equation}
E_n \sim C \, (Z^{\ast})^n \, n^b \, e^{-a\sqrt{n}} \left( 1 + \frac{\gamma_1}{n^{1/2}} + \frac{\gamma_2}{n} + ... \right) \label{fitbaker}
\end{equation}
Ivanov and Dubau showed that the leading term of this asymptotic law can be reproduced if one assumes that the behavior of the imaginary part of the energy near the point $Z^{\ast}$ (for $Z < Z^{\ast}$) is given by:
\begin{equation}
\mbox{Im}(E) \sim A \left( 1 - \frac{Z}{Z^{\ast}} \right)^p \exp \left( -\frac{c}{1-\frac{Z}{Z^{\ast}}} \right), \label{fit}
\end{equation}
where
\begin{equation}
p = -2 b - \frac{3}{2} \mbox{  and  } c = \frac{a^2}{4}. \label{fitparam}
\end{equation}
We performed a least-squares fit of the coefficients $E_n$ from~\cite{baker1990} in the range $n \in [101,401]$ by the first term in expression~(\ref{fitbaker}) (more precisely, we fitted $\ln (E_n)$ by the logarithm of that term with equal weights for all data points); $Z^{\ast}$ was set equal to $0.911 \, 028 \, 22$. We find
\begin{equation}
a = 0.26374(5) \, \mbox{,} \;\; b = -1.9896(4). \label{myfit}
\end{equation}
These values are close to those obtained in~\cite{baker1990} using a different procedure,
\begin{equation}
a \approx 0.272 \, \mbox{,} \;\; b \approx -1.94. \label{bakersfit}
\end{equation}
Then we performed a least-squares fit of $\mbox{Im} [E(Z)]$ by~(\ref{fit}) in the range $Z \in [0.905,0.9103]$ (again, we fit $\ln(\mbox{Im} [E(Z)])$ by the logarithm of~(\ref{fit})). Here, $Z^{\ast}$ was taken as a fitting parameter. This yields
\begin{equation}
p = 2.42(9) \, \mbox{,} \; c = 0.0180(5) \;\; \mbox{and} \;\; Z^{\ast} = 0.911 \, 276(12).
\end{equation}
The values of $p$ and $c$ are in very good agreement with those deduced from~(\ref{fitparam}) and (\ref{myfit}), $p = 2.48$ and $c=0.0174$ (or $p = 2.38$ and $c=0.0185$ if one adopts the values of $a$ and $b$ from~(\ref{bakersfit}) reported in~\cite{baker1990}). The value of $Z^{\ast}$ deduced from this fit is slightly too large, by $2.5 \, 10^{-4}$ ($Z^{\ast}$ cannot be larger than $Z_c$). We also tried fitting by a refined version of~(\ref{fit}) which was derived in Ref.~\cite{dubau1998} to reproduce also the next-to-leading-order term of~(\ref{fitbaker}), but this did not improve the agreement. This discrepancy could be related to the numerical difficulties discovered by Baker et al.~\cite{baker1990}: the asymptotic behavior of the perturbation coefficients $E_n$ manifests itself only at very high orders, and the nature of the singularity of $E(Z)$ is revealed only extremely close to $Z = Z^{\ast}$. It is thus possible that a more detailed investigation of the asymptotic behavior of $E_n$ could lead to a more accurate model of the variation of ${\rm Im}(E)$ vs. $Z$.

Despite the discrepancy on $Z^{\ast}$, the good quality of the fit suggests that the expression~(\ref{fit}) accounts for the essential features of the variation of ${\rm Im}(E)$. In particular, it shows that ${\rm Im}(E)$ does not drop to zero immediately beyond the last investigated value at $Z = 0.9103$ but continues to have a nonzero values up to a close vicinity of $Z_c$. This strengthens the hypothesis of the ground state acquiring a finite width exactly at $Z = Z_c$, implying the identity $Z^{\ast} = Z_c$.

\section{Conclusion}

We have studied the atomic state in two-electron atoms slightly below the critical nuclear charge $Z_c = \numprint{0.911028}...$ by variational calculations combined with the complex rotation method. It was shown that already in a close vicinity of $Z = Z_c$, the bound state is transformed into a resonance whose width increases very rapidly as $Z$ is decreased further away from $Z_c$. The variation of the width as a function of $Z$ is in good agreement with a model derived from the analysis of the $1/Z$ perturbation series, suggesting that the state acquires a finite width exactly at $Z = Z_c$. It would be interesting to get an independent confirmation by a recalculation and analysis of the $1/Z$ perturbation series, up to even higher orders than what was done in~\cite{baker1990}.

\vspace{3mm}

\textbf{Acknowledgments.} I thank A.V. Turbiner for bringing the critical charge problem to my attention and for stimulating discussions. I would also like to thank D. Delande and L. Hilico for helpful dicsussions and reading of the manuscript. I acknowledge support as a fellow of the Institut Universitaire de France.

\end{document}